\documentclass[
 aip,
 amsmath,amssymb,
 reprint,
]{revtex4-1}

\usepackage{graphicx}
\usepackage{dcolumn}
\usepackage{bm}

\usepackage[utf8]{inputenc}
\usepackage[T1]{fontenc}
\usepackage{mathptmx}
\usepackage{etoolbox}

\usepackage{bm,ascmac,amsmath,amssymb,amsthm,mathrsfs,amsfonts,dsfont}
\usepackage{epsfig}
\usepackage{braket}
\usepackage{enumerate}
\usepackage{xcolor}
\usepackage{float}
\usepackage{algorithm}
\usepackage{algorithmicx}
\usepackage{algpseudocode}
\usepackage{comment}
\usepackage{appendix}
\usepackage{here}
\usepackage{tabularx}
\usepackage{dcolumn}
\usepackage[caption=false]{subfig}
\usepackage{adjustbox}
\usepackage[hidelinks]{hyperref}
\usepackage{bookmark}

\theoremstyle{definition}

\usepackage{multirow}
\usepackage{url}
\usepackage{braket}
\usepackage{tikz}
\usepackage{float}

\makeatletter
\def\@email#1#2{
 \endgroup
 \patchcmd{\titleblock@produce}
  {\frontmatter@RRAPformat}
  {\frontmatter@RRAPformat{\produce@RRAP{*#1\href{mailto:#2}{#2}}}\frontmatter@RRAPformat}
  {}{}
}
\makeatother

\begin{document}

\newfloat{protocol}{htbp}{idf}
\floatname{protocol}{Protocol~}

\newfloat{resource}{htbp}{idf}
\floatname{resource}{Resource~}

\newfloat{simulator}{htbp}{idf}
\floatname{simulator}{Simulator~}

\newfloat{problem}{htbp}{idf}
\floatname{problem}{Problem~}

\preprint{APS/123-QED}

\title
[Quantum simulation of many-body dynamics with noise-robust Trotter decomposition based on symmetric structures]
{Quantum simulation of many-body dynamics with noise-robust Trotter decomposition based on symmetric structures}

\author{Bo Yang}
\email{Bo.Yang@lip6.fr}
\thanks{
    The authors contributed equally.
}
\affiliation{LIP6, Sorbonne Université, CNRS, 4 place Jussieu, 75005 Paris, France}
\affiliation{Graduate School of Information Science and Technology, The University of Tokyo, Bunkyo-ku, Tokyo 113-8656, Japan}

\author{Naoki Negishi}
\email{negishi@roma2.infn.it}
\thanks{
    The authors contributed equally.
}
\affiliation{Graduate School of Arts and Sciences, The University of Tokyo, Meguro-ku, Tokyo, 153-8902, Japan}
\affiliation{Dipartimento di Fisica, Università di Roma Tor Vergata, Via della Ricerca Scientifica 1, 00133 Rome, Italy}
\affiliation{INFN, Sezione di Roma Tor Vergata, Via della Ricerca Scientifica 1, 00133 Rome, Italy}

\date{\today}

\begin{abstract}
The Suzuki-Trotter decomposition, which digitalizes quantum time evolution, provides a promising framework for simulating quantum dynamics on quantum hardware and exploring quantum advantage over classical computation.
However, conventional Trotter circuits require a large number of non-local gates, lowering their faithfulness to the ideal dynamics when implemented on current noisy quantum hardware.
While most previous studies have focused on circuit optimization, we instead propose a new Trotter decomposition that is intrinsically circuit-efficient for simulating quantum dynamics on near-term devices.
Our method substantially reduces both the residual error by Trotter decomposition and the number of CNOT operations compared to conventional Trotter decompositions by exploiting the symmetry of the target model to construct an effective Hamiltonian with fewer two-qubit gates.
We demonstrate the noise robustness of the proposed approach through numerical simulations of a nine-site Heisenberg model under realistic noise, and further validate its experimental practicality on the IBM superconducting device, achieving a state fidelity exceeding $0.98$ when combined with quantum error mitigation in the three-site case.
The proposed circuit design is also compatible with existing circuit optimization techniques.
Our results establish a practical route toward noise-resilient quantum simulation in many-body dynamics.
\end{abstract}

\maketitle

\section{\label{sec:introduction}Introduction}

Through the rapid advance of quantum hardware, quantum simulation has gained emerging attention to simulate physical and chemical models that become practically intractable by classical computational resources~\cite{Lanes2025A, Huang2025The}.
Its prominent and versatile applicability lies in non-equilibrium quantum many-body dynamics, where the digital simulation based on Suzuki-Trotter decomposition enables tractable and scalable approximation of real-time evolution~\cite{Trotter1959On, Suzuki1976Generalized, Hatano2005Finding}.
Simulating Trotter iterations with quantum computers frees from classical simulation with exponential computational resources, requiring only linear overhead to system size, which is thus seen as one of the applications with potential near-term quantum advantage.

However, the non-negligible noise level and hardware restrictions of current quantum hardware still pose a significant obstacle to the practical realization of such Trotter-based simulations.
In particular, superconducting quantum devices~\cite{ibm_quantum}, which are among the most extensively developed and commercially accessible platforms, suffer from noisy non-local gates and limited coherence times~\cite{mooney2021generation, Yang2022Testing}.
Therefore, it is essential to design quantum circuits with reduced depth and fewer non-local gates, such as CNOT gates, to alleviate noise accumulation and improve the fidelity of simulations.

While substantial efforts have focused on optimizing given Trotter circuits under hardware constraints~\cite{Lotstedt2023Error-mitigated, Chowdhury2024Enhancing, Choi2025Quantum,YangPRA2022,ZhaoPRXQ2023,Jones2019ACM}, the underlying Trotter decomposition itself sets the fundamental limits of such optimization.
In this work, we design a new alternative Trotter decomposition strategy that substantially reduces both the residual error by Trotter decomposition and the number of CNOT gates in use.
To achieve this, we exploit the symmetric structure of the given Hamiltonian, particularly, of the $XXX$ Heisenberg model.
In particular, we transform the three-site Heisenberg Hamiltonian into a more concise two-site effective Hamiltonian through an encoding and decoding procedure.
This transformation enables a faster convergence rate of the Trotter iterations compared with using the conventional Trotter decomposition, thereby requiring fewer Trotter iterations to achieve the same level of residual error.

\begin{figure}
    \centering
    \includegraphics[width=\linewidth]{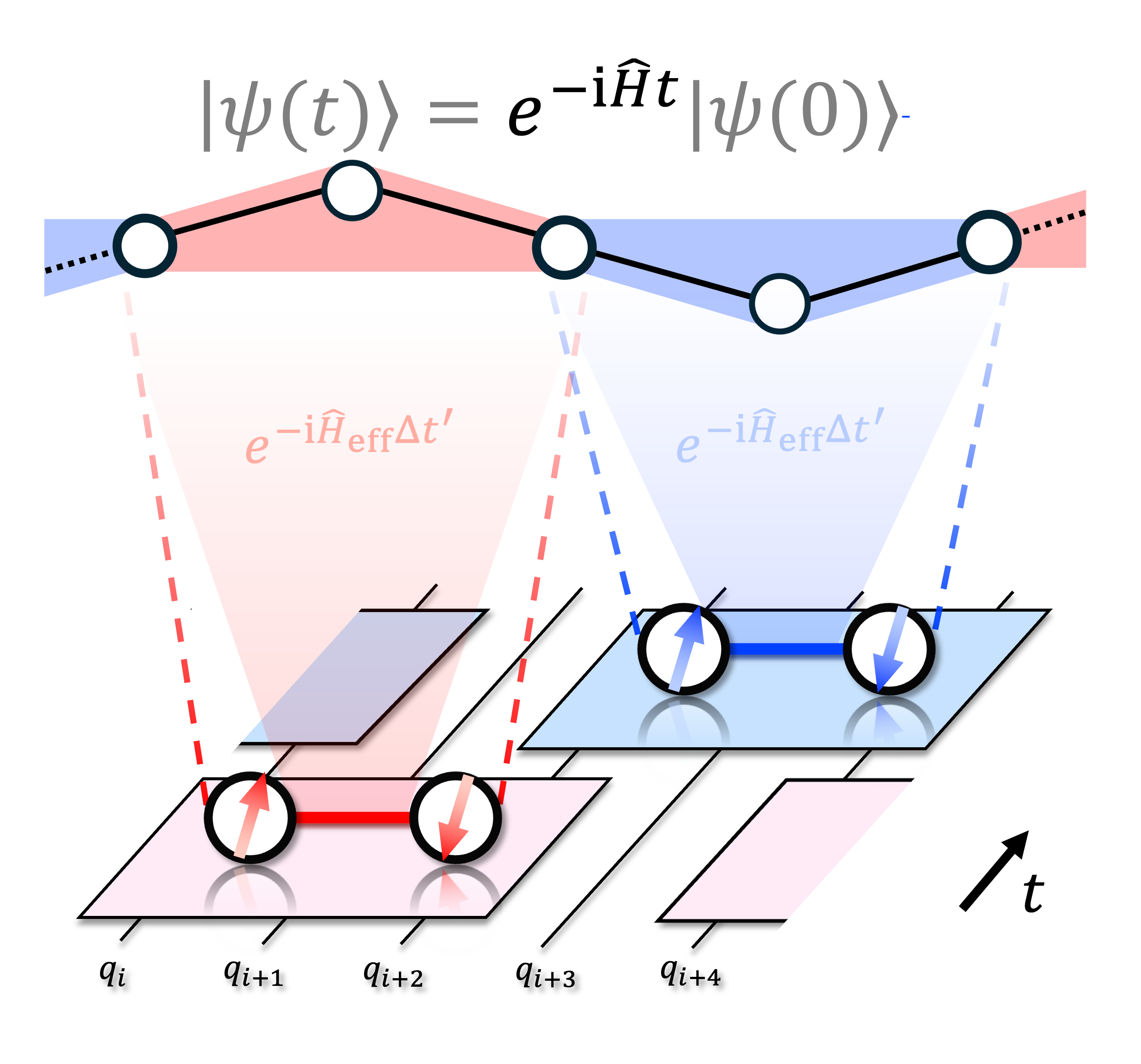}
    \caption{
        The schematic illustration of the proposed framework.
        In simulating the time evolution of a given Hamiltonian, we exploit its symmetric structure to construct Trotter blocks with a faster convergence rate and fewer CNOT gates based on the effective Hamiltonian.
    }
    \label{fig:highlight}
\end{figure}

This new decomposition with the effective Hamiltonian reduces the average number of CNOT gates in each Trotter step to $1.75$ per qubit, whereas the conventional Trotter circuit requires $3$.
The proposed method highlights the potential of reducing the circuit overhead with a more efficient approach for Trotter decomposition rather than merely performing circuit optimization on existing Trotter circuits.
The schematic illustration of the proposed method can be found in Fig.~\ref{fig:highlight}.

We demonstrate the noise-robustness of our proposed method through numerical simulation.
Our method outperforms the conventional Trotter decomposition in simulating the time evolution of the nine-site $XXX$ Heisenberg model in both setups without noise and with depolarizing noise.
We also simulate the time evolution of a three-site $XXX$ Heisenberg model on the superconducting quantum device \verb+ibmq_jakarta+ provided by the IBM Quantum Platform~\cite{ibm_quantum}.
Using quantum error mitigation (QEM)~\cite{li2017efficient, Temme_2017, endo2018practical, digital_zne, yang2022efficient, Yang2025Resource-Efficient, cai2023quantum, endo2021hybrid}, we achieve the target state fidelity over $0.98$ on \verb+ibmq_jakarta+.
In implementing the proposed Trotter circuits, our method finds further compatibility with circuit optimization with the Qiskit package~\cite{qiskit2024}.
The overall experiments demonstrate that our method offers not only a novel, efficient Trotter decomposition scheme but also a practical and feasible solution for simulating physical models on current quantum hardware.

\section{\label{sec:Suzuki-Trotter Decomposition}Suzuki-Trotter Decomposition}

We consider $N$-site $J=1$ $XXX$ Heisenberg Hamiltonian with $N=2M+1$, $M\in \mathbb{Z}_{\ge0}$, and open boundary condition formalized as
\begin{equation}
    \hat{H}=\sum_{i=1}^{N-1}\vec{\sigma}^{(i)}\cdot\vec{\sigma}^{(i+1)},\label{eq:def_HHeis}
\end{equation}
where $\cdot$ denotes the inner product of three components of the Pauli operators of $i$-th site, $\hat{\sigma}_x^{(i)}:=\begin{pmatrix}
0&1\\1&0\end{pmatrix}$,$\hat{\sigma}_y^{(i)}:=\begin{pmatrix}
0&-\mathrm{i}\\\mathrm{i}&0\end{pmatrix}$, and $\hat{\sigma}_z^{(i)}:=\begin{pmatrix}
1&0\\0&-1\end{pmatrix}$, defined as
\begin{align}
    \vec{\sigma}^{(i)}\cdot\vec{\sigma}^{(j)}:=\sum_{\mu\in\{x,y,z\}}\hat{\sigma}_\mu^{(i)}\otimes\hat{\sigma}^{(j)}_\mu \quad \text{for}\quad i\not=j\label{eq:def_{i}npro_pauli}.
\end{align}

\begin{figure}[htbp]
    \centering
    \includegraphics[width=\linewidth]{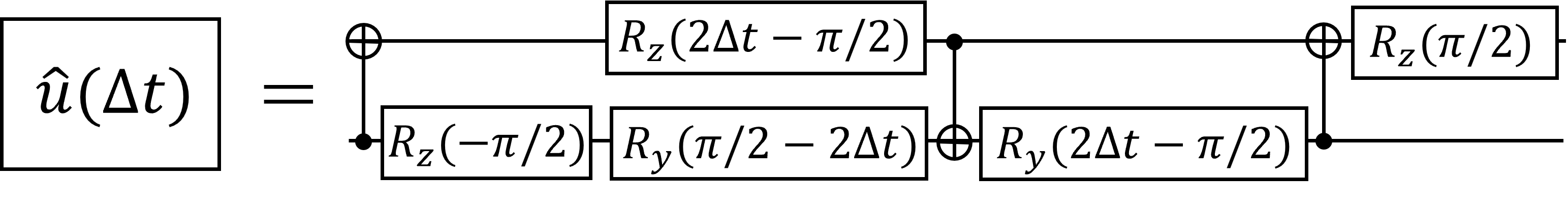}
    \caption{
        The quantum circuit of the unitary operator in Eq.~\eqref{eq:unitprop}.
    }
    \label{fig:Unitprop}
\end{figure}

Given a Heisenberg Hamiltonian $\hat{H}=\hat{O}_{1}+\hat{O}_{2}$ that consists of two non-commutative operators $\hat{O}_{1}$ and $\hat{O}_{2}$, i.e. $[\hat{O}_{1},\hat{O}_{2}]\not=0$, the conventional Trotter decomposition with $n$ steps~\cite{Trotter1959On, Suzuki1976Generalized, Hatano2005Finding} approximates the evolution of this Hamiltonian with 
\begin{equation}
\begin{split}
    \hat{U}(t)
    &=\biggl(\exp\left(-{\mathrm{i}}\hat{O}_{2}\Delta t\right)\exp\left(-{\mathrm{i}}\hat{O}_{1}\Delta t\right)\biggr)^{n} \\
    &\quad + \mathcal{O}\left(\hat{\epsilon}_{1}n^{-1}\right),\label{eq:Suzuki_Trotter}
\end{split}
\end{equation}
where $\Delta t=t/n$ denotes the evolution time for a step, and $\hat{O}_{1}$ and $\hat{O}_{2}$ are chosen to be
\begin{align}
    \hat{O}_{1}&=\sum_{i=1}^{M}\vec{\sigma}^{(2i-1)}\cdot\vec{\sigma}^{(2i)}\label{eq:def_O1}, \\
    \hat{O}_{2}&=\sum_{i=1}^{M}\vec{\sigma}^{(2i)}\cdot\vec{\sigma}^{(2i+1)}\label{eq:def_O2},
\end{align}
and $\hat{\epsilon}_{1}$ is the operator in the error term as follows:
\begin{align}
    \hat{\epsilon}_{1}&=[\hat{O}_{1},\hat{O}_{2}]\notag\\
    &=\sum_{i=2}^{N-1}(-1)^{i}[\vec{\sigma}^{(i-1)}\cdot\vec{\sigma}^{(i)},\vec{\sigma}^{(i)}\cdot\vec{\sigma}^{(i+1)}].
\end{align}
The quantum circuit for the decomposition Eq.~\eqref{eq:Suzuki_Trotter} can be constructed in the following way.
First, let $\hat{u}^{(i)}(\Delta t)$ be the unitary operator of evolution time $\Delta t$ for $i$-th and $i+1$-th qubit defined as
\begin{align}\label{eq:unitprop}
    \hat{u}^{(i)}(\Delta t)=\exp\left(-{\mathrm{i}}\vec{\sigma}^{(i)}\cdot\vec{\sigma}^{(i+1)}\Delta t\right).
\end{align}
Since any two-qubit unitary operation can be realized by three CNOT gates~\cite{Vatan2004Optimal, Vidal2004Universal, Shende2006Synthesis}, Fig.~\ref{fig:Unitprop} provides a quantum circuit to implement $\hat{u}^{(i)}(t)$.

\begin{figure}[H]
    \centering
    \includegraphics[width=0.8\linewidth]{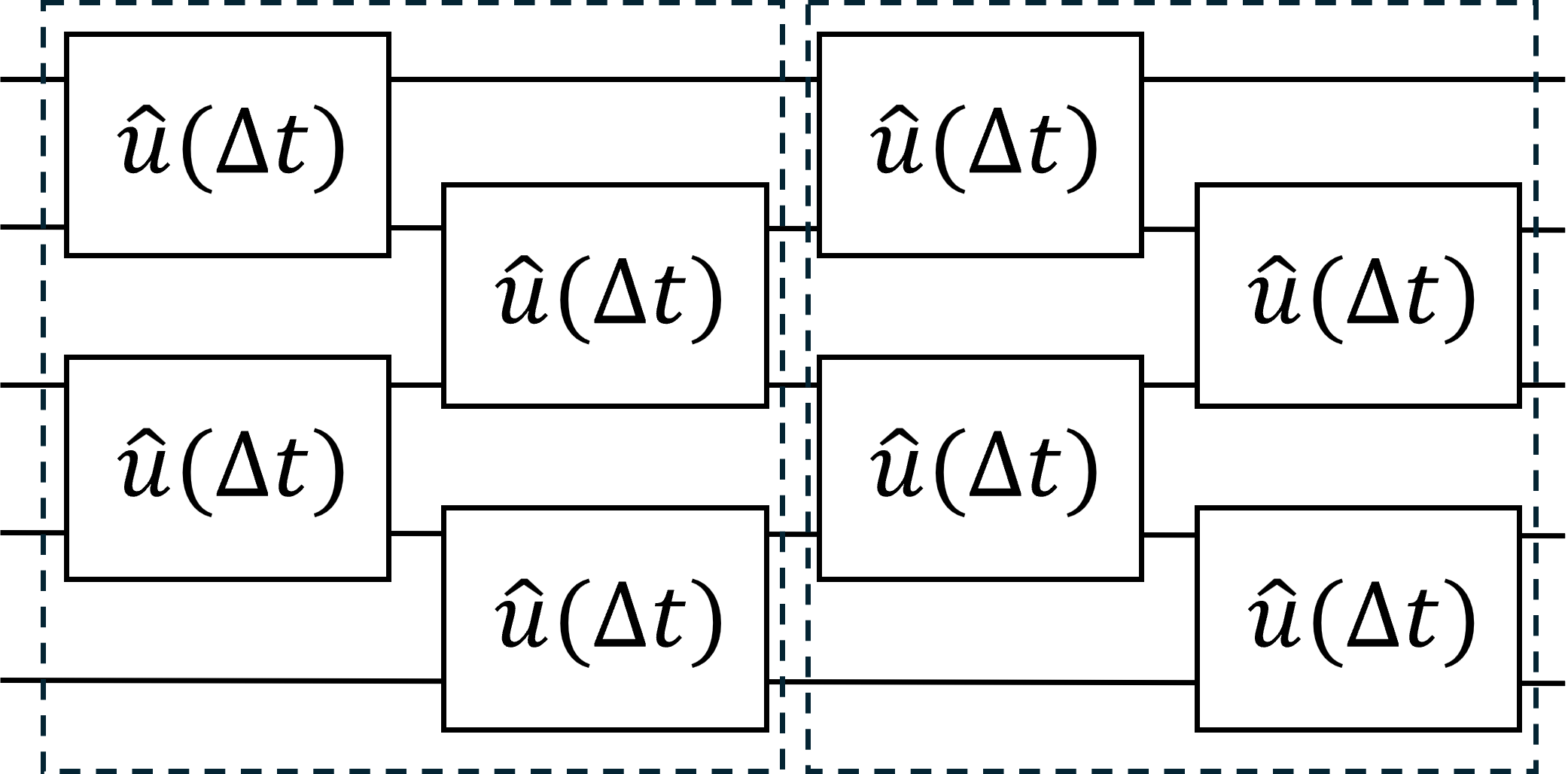}
    \caption{
        The quantum circuit of the time evolution operator using the Trotter decomposition in Eq.~\eqref{eq:Suzuki_Trotter} for the $5$-site system.
        The operation corresponding to the propagation of each single time step $\Delta t$ is enclosed by a dashed box.
    }
    \label{fig:OldTrotter}
\end{figure}

Using this circuit block, the Trotter decomposition in Eq.~\eqref{eq:Suzuki_Trotter} is then constructed by the quantum circuit in Fig.~\ref{fig:OldTrotter}.
Since each time step has two layers of operator $\hat{u}^{(i)}(t)$, the averaged number of CNOT gates applied over each qubit is three, which represents the most efficient circuit construction regarding the CNOT overhead known to date~\cite{PRR2024Talal}.

\section{\label{sec:Proposed Decomposition}Proposed Decomposition}

\begin{figure}[htbp]
    \centering
    \subfloat[]{
        \includegraphics[width=\linewidth]{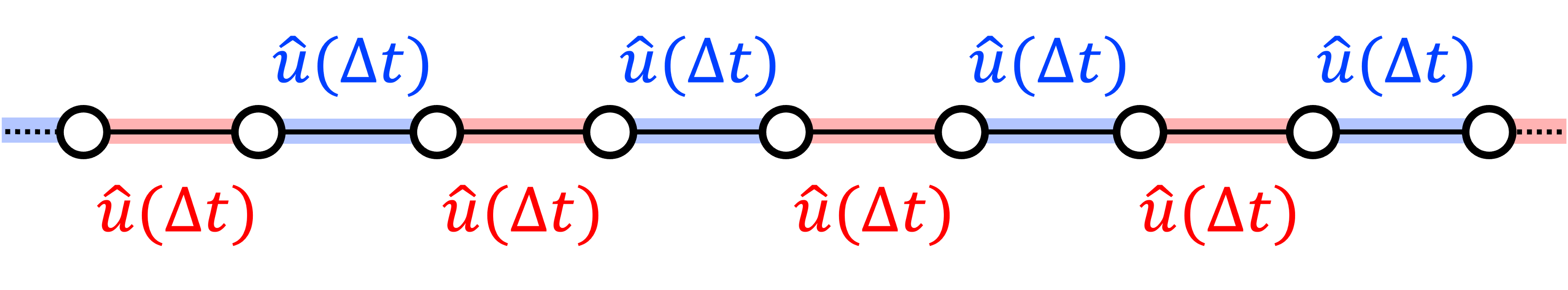}
    }
    \hfill
    \subfloat[]{
        \includegraphics[width=\linewidth]{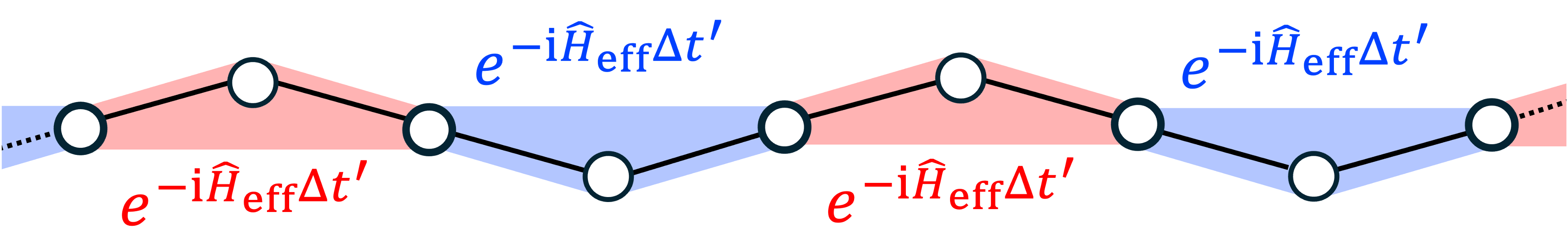}
    }
    \caption{
        (a) The schematic illustration of the conventional Trotter decomposition that takes non-commutative Trotter blocks with propagator $\hat{u}(\Delta t)$ according to the native edge structure of the given Hamiltonian.
        (b) The schematic illustration of the proposed Trotter decomposition that takes non-commutative Trotter blocks with propagator $\hat{U}_{\mathrm{enc}}^{\dagger}\exp\left(-\mathrm{i}\hat{H}_{\mathrm{eff}}\Delta t^{\prime}\right)\hat{U}_{\mathrm{enc}}$ according to the hyper-edge of the set with three neighboring vertices in the given Hamiltonian.
    }
    \label{fig:concept}
\end{figure}

In order to reduce the number of CNOT gates in each single time step, we propose a new Trotter decomposition.
First, we partition the Heisenberg Hamiltonian into $\hat{H}=\hat{A}_{1}+\hat{B}_{1}$ with the operators
\begin{align}
    \hat{A}_{1}&=\sum_{i=1}^{M^{\prime}}\vec{\sigma}^{(4i-3)}\cdot\vec{\sigma}^{(4i-2)}+\vec{\sigma}^{(4i-2)}\cdot\vec{\sigma}^{(4i-1)},\label{eq:A1} \\
    \hat{B}_{1}&=\sum_{i=1}^{M^{\prime\prime}}\vec{\sigma}^{(4i-1)}\cdot\vec{\sigma}^{(4i)}+\vec{\sigma}^{(4i)}\cdot\vec{\sigma}^{(4i+1)},\label{eq:B1}
\end{align}
where $M^{\prime}=M^{\prime\prime}=M/2$ for $M\in 2\mathbb{Z}$ and $M^{\prime}=M^{\prime\prime}+1=(M+1)/2$ for $M \not\in 2\mathbb{Z}$.
Likewise, we also propose another decomposition $\hat{H}=\hat{A}_{2}+\hat{B}_{2}$ with the operators
\begin{align}
    \hat{A}_{2}&=\delta_{M,2N^{\prime}+1}\vec{\sigma}^{(2M)}\cdot\vec{\sigma}^{(2M+1)}\notag\\
    &+\sum_{i=1}^{N^{\prime}}\vec{\sigma}^{(4i-2)}\cdot\vec{\sigma}^{(4i-1)}+\vec{\sigma}^{(4i-1)}\cdot\vec{\sigma}^{(4i)},\label{eq:A2} \\
    \hat{B}_{2}&=\vec{\sigma}^{(1)}\cdot\vec{\sigma}^{(2)}+\delta_{M,2N^{\prime}}\vec{\sigma}^{(2M)}\cdot\vec{\sigma}^{(2M+1)}\notag\\
    &+\sum_{i=1}^{N^{\prime\prime}-1}\vec{\sigma}^{(4i)}\cdot\vec{\sigma}^{(4i+1)}+\vec{\sigma}^{(4i+1)}\cdot\vec{\sigma}^{(4i+2)},\label{eq:B2}
\end{align}
where $N^{\prime}=N^{\prime\prime}=M/2$ for $M\in 2\mathbb{Z}$ and $N^{\prime}=N^{\prime\prime}-1=(M-1)/2$ for $M \not\in 2\mathbb{Z}$.

Using $\hat{A}_{k}$ and $\hat{B}_{k}$, we define a single Trotter iteration as a sequence of propagators
\begin{equation}
    \exp\left(-{\mathrm{i}}\hat{B}_{2}\Delta t^{\prime}\right)
        \exp\left(-{\mathrm{i}}\hat{A}_{2}\Delta t^{\prime}\right)\exp\left(-{\mathrm{i}}\hat{B}_{1}\Delta t^{\prime}\right)
        \exp\left(-{\mathrm{i}}\hat{A}_{1}\Delta t^{\prime}\right).
\end{equation}
Then, the unitary operation that evolves the system for the time $t$ using $m$ proposed Trotter iterations with $\displaystyle \Delta t^{\prime}=t/(2m)$ is described as
\begin{equation}
\begin{split}
    \hat{U}(t)
    &=\Biggl(
        \exp\left(-{\mathrm{i}}\hat{B}_{2}\Delta t^{\prime}\right)
        \exp\left(-{\mathrm{i}}\hat{A}_{2}\Delta t^{\prime}\right) \\
        &\quad\quad\quad \exp\left(-{\mathrm{i}}\hat{B}_{1}\Delta t^{\prime}\right)
        \exp\left(-{\mathrm{i}}\hat{A}_{1}\Delta t^{\prime}\right)
    \Biggr)^{m} \\
    &\quad+\mathcal{O}\left(\hat{\epsilon}_{2}m^{-1}\right),
    \label{eq:New_Trotter}
\end{split}
\end{equation}
with an operator $\hat{\epsilon}_{2}$ defined as
\begin{align}
    \hat{\epsilon}_{2}
    &=\frac{1}{4}[\hat{A}_{1},\hat{B}_{1}]+\frac{1}{4}[\hat{A}_{2},\hat{B}_{2}]\notag\\
%    &=\frac{1}{4}\sum_{j=2}^{N-1}\sqrt{2}\sin\left(\frac{\pi}{4}(2j-1)\right)[\vec{\sigma}^{(j-1)}\cdot\vec{\sigma}^{(j)},\vec{\sigma}^{(j)}\cdot\vec{\sigma}^{(j+1)}]
&=\frac{1}{4}\sum_{j=1}^{M}(-1)^j[\vec{\sigma}^{(2j-1)}\cdot\vec{\sigma}^{(2j)},\vec{\sigma}^{(2j)}\cdot\vec{\sigma}^{(2j+1)}]\notag\\
&\quad+\frac{1}{4}\sum_{j=1}^{M-1}(-1)^j[\vec{\sigma}^{(2j)}\cdot\vec{\sigma}^{(2j+1)},\vec{\sigma}^{(2j+1)}\cdot\vec{\sigma}^{(2j+2)}],
\label{eq:New_Trotter_error}
\end{align}

Remarkably, the number of required Trotter iterations $m$ in Eq.~\eqref{eq:New_Trotter} is smaller than $n$ in Eq.~\eqref{eq:Suzuki_Trotter} to achieve the same level of residual error by Trotter decomposition.
This arises from the fact that, as shown in Fig.~\ref{fig:concept}, the proposed Trotter iteration has sparser decomposition interval than the conventional one, which yields fewer combinations of non-commutative operators that contributes to the residual error.
In particular, the residual error of Eqs.~\eqref{eq:New_Trotter} and ~\eqref{eq:New_Trotter_error} is reduced to one-quarters of that of the conventional Trotter decomposition for $m=n$. 
This implies that using $m=n/4$ iterations of the proposed Trotter blocks achieves the same level of residual error as the conventional Trotter blocks.
%In designing a quantum circuit for this new Trotter decomposition, we can use $\{\hat{u}^{(4i-1)}(\Delta t^{\prime})\}_{i=1}^{M}$ and $\{\hat{u}^{(4i)}(\Delta t^{\prime})\}_{i=1}^{M}$ to construct the time evolution operators $\exp\left(-{\mathrm{i}}\hat{O}_{1}'\Delta t^{\prime})$ and $\exp\left(-{\mathrm{i}}\hat{O}_{2}'\Delta t^{\prime})$ in Eq.~\eqref{eq:New_Trotter} since they share the same form as Eq.~\eqref{eq:unitprop}.

Next, we aim at designing efficient quantum circuits of the time evolution operator $\exp\left(-{\mathrm{i}}\hat{A}_{k}\Delta t^{\prime}\right)$ and $\exp\left(-{\mathrm{i}}\hat{B}_{k}\Delta t^{\prime}\right)$, where $\hat{A}_{k}$ and $\hat{B}_{k}$ take the form of a three-site $XXX$ Heisenberg Hamiltonian $\hat{H}_{3}=\vec{\sigma}^{(1)}\cdot\vec{\sigma}^{(2)}+\vec{\sigma}^{(2)}\cdot\vec{\sigma}^{(3)}$.
To construct an efficient quantum circuit of $\exp\left(-{\mathrm i}\hat{H}_{3}t\right)$, we compress it to a smaller subsystem by deriving an effective Hamiltonian $H_{\mathrm{eff}}$ focusing on the $\mathrm{SU}(2)$ symmetry that $\hat{H}_{3}$ is equipped with.
We use the fact that the three-site Hamiltonian $\hat{H}_{3}$ commutes with the operator defined as $\hat{s}_{\mu}:=-\hat{\sigma}_\mu^{(1)}\otimes\hat{\sigma}_\mu^{(2)}\otimes\hat{\sigma}_\mu^{(3)}$, i.e. $[\hat{H}_{3}, \hat{s}_{\mu}]=0$ and that $\mathrm{SU}(2)$ group is spanned by $\{\hat{s}_{\mu}\}_{\mu=x,y,z}$.
This symmetry structure yields simultaneous eigenstates as follows:
\begin{align}
    \hat{H}_{3}\ket{E}\otimes\ket{P}
    &= E\ket{E}\otimes\ket{P}, \label{eq:indepSE} \\    
    \hat{s}_{z}\ket{E}\otimes\ket{P}
    &= P\ket{E}\otimes\ket{P}, \label{eq:Parity}
\end{align}
where $\ket{E}$ and $\ket{P}$ denote the eigenstate of the energy $E$ and the eigenvalue $P\in\{-1,1\}$ of $\hat{s}_{z}$, respectively.
This implies that the eigenstates are doubly degenerated regarding the energy $E$.

\begin{figure}[htbp]
    \centering
    \includegraphics[width=0.4\linewidth]{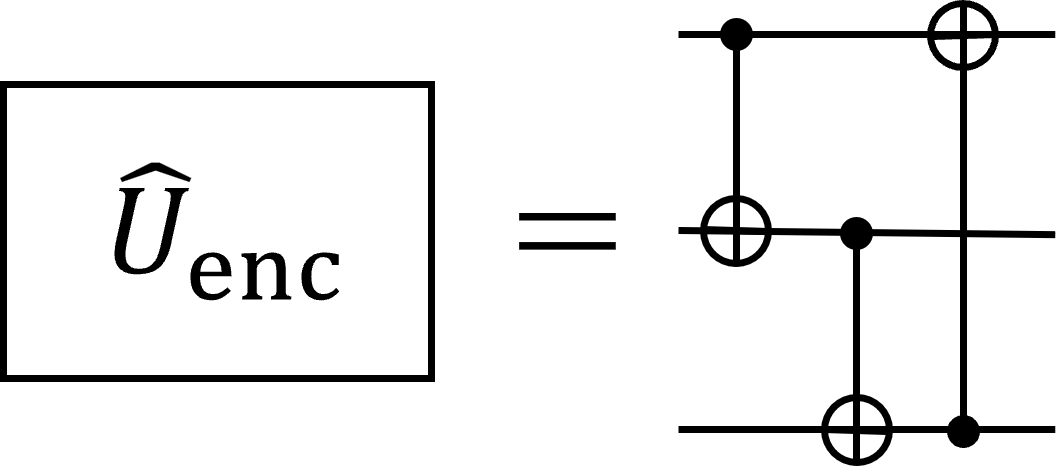}
    \caption{
        The quantum circuit to realize the encoder $\hat{U}_{\mathrm{enc}}$.
    }
    \label{fig:Uenc}
\end{figure}
This double degeneracy stemming from $\mathrm{SU}(2)$ symmetry is essential to compress $\hat{H}_3$ into a lower dimensional effective Hamiltonian $\hat{H}_{\rm eff}$.
Thanks to this degeneracy, one can encode the three-site state into the composition of a single-qubit system specifying the eigenvalue $P$ and the remaining two-qubit system specifying the state within the subspace corresponding to $P$.
This encoding is represented by a unitary $\hat{U}_{\mathrm{enc}}$ that transforms the basis $\ket{E}\otimes\ket{P}$ into another separable state,
\begin{align}
    \hat{U}_{\mathrm{enc}}\ket{E}\otimes\ket{1}&=\ket{0}\otimes\ket{\Psi_{E}}, \label{eq:Uenc1}\\
    \hat{U}_{\mathrm{enc}}\ket{E}\otimes\ket{-1}&=\ket{1}\otimes\ket{\Psi_{E}}, \label{eq:Uenc2}
\end{align}
where $\{\ket{0},\ket{1}\}$ denotes the state in the single-qubit system and $\ket{\Psi_{E}}$ denotes the state in the two-qubit system.
This encoder $\hat{U}_{\mathrm{enc}}$ can be constructed with three CNOT gates, as shown in Fig.~\ref{fig:Uenc}.

\begin{figure}[htbp]
    \centering
    \includegraphics[width=\linewidth]{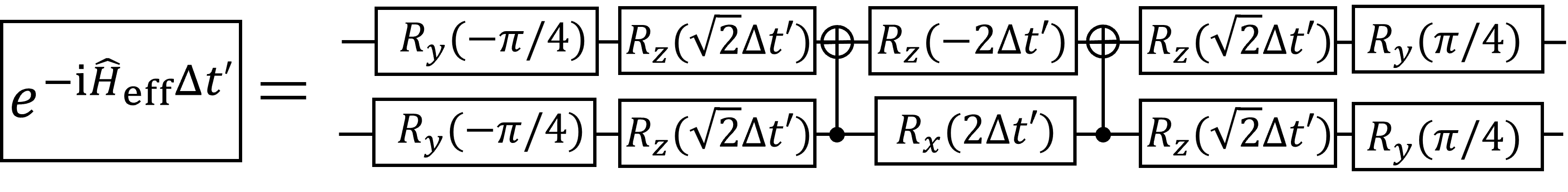}
    \caption{
        The quantum circuit of the time evolution operator $\exp\left(-{\mathrm{i}}\hat{H}_{\mathrm{eff}}\Delta t^{\prime}\right)$ given by Eq.~\eqref{eq:propeff}.
    }
    \label{fig:Trotter_unit}
\end{figure}

The encoder $\hat{U}_{\mathrm{enc}}$ then transforms $\hat{H}_{3}$ to a two-qubit effective Hamiltonian $\hat{H}_{\mathrm{eff}}$,
\begin{equation}
\begin{split}
    \hat{H}_{\mathrm{eff}}
    &=\hat{U}_{\mathrm{enc}}\hat{H}_{3}\hat{U}_{\mathrm{enc}}^{\dagger}\\
    &=\sqrt{2}(\hat{h}^{(1)}+\hat{h}^{(2)})-(\hat{\sigma}_z^{(1)}\otimes\hat{\sigma}_x^{(2)}+\hat{\sigma}_x^{(1)}\otimes\hat{\sigma}_z^{(2)}). \label{eq:Heff_xz}
\end{split}
\end{equation}
where $\hat{h}^{(i)}=(\hat{\sigma}_x^{(i)}+\hat{\sigma}_z^{(i)})/\sqrt{2}$ is the Hadamard operator.
This yields the following equivalence between the time evolution operators:
\begin{equation}
\begin{split}
    \exp\left(-{\mathrm{i}}\hat{H}_{3}\Delta t^{\prime}\right)
    &= \hat{U}_{\mathrm{enc}}^{\dagger}\exp\left(-{\mathrm{i}}\hat{H}_{\mathrm{eff}}\Delta t^{\prime}\right)\hat{U}_{\mathrm{enc}}. \label{eq:Heff_time_evolve}
\end{split}
\end{equation}
This suggests that the time evolution under $\hat{H}_{3}$ can be realized as the composition of the encoder $\hat{U}_{\mathrm{enc}}$, the time evolution under $\hat{H}_{\mathrm{eff}}$, and the decoding unitary $\hat{U}_{\mathrm{enc}}^{\dagger}$.

\begin{figure}
    \centering
    \includegraphics[width=\linewidth]{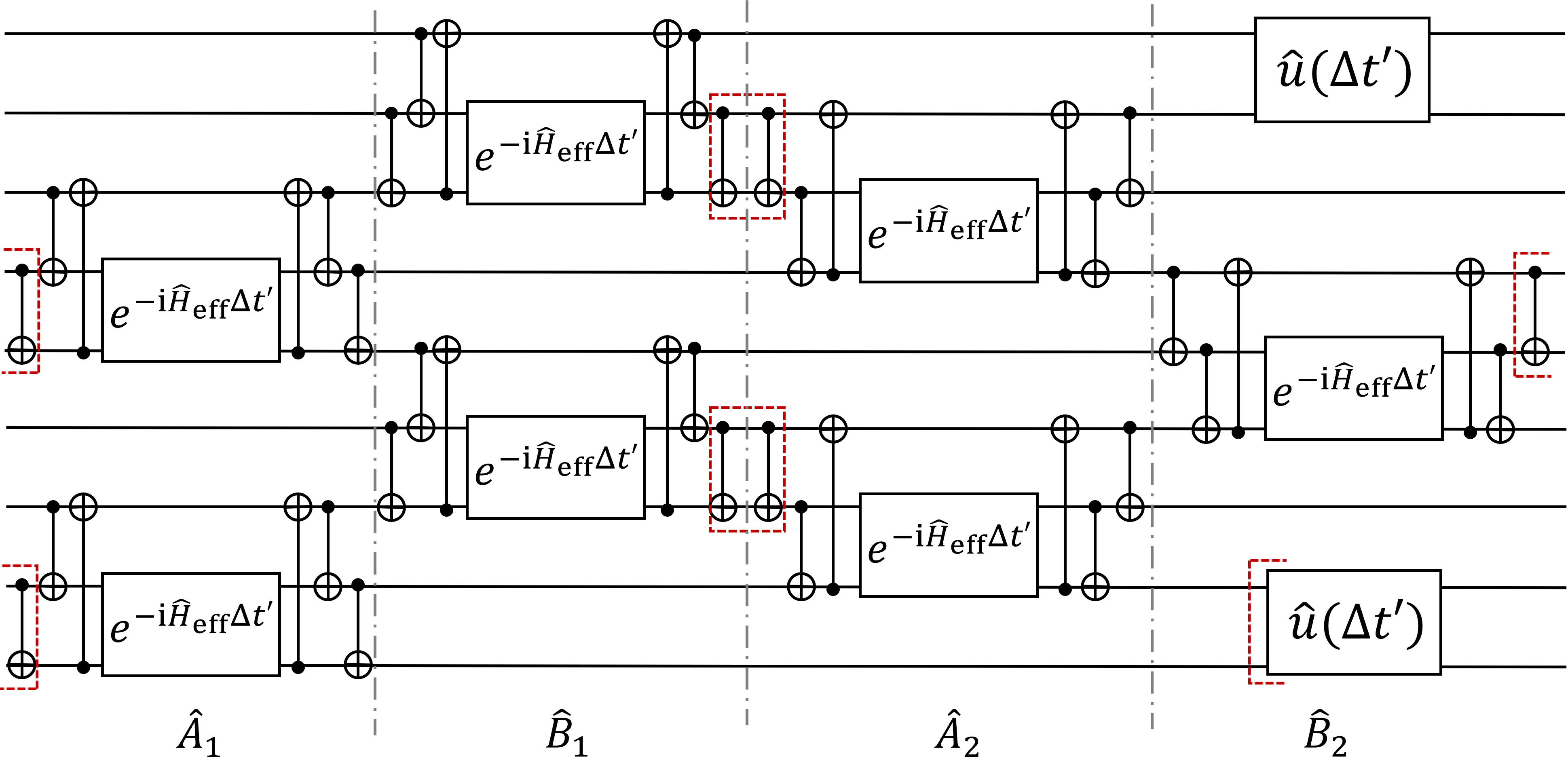}
    \caption{
        The quantum circuit of the single Trotter block for the $9$-site system, denoting $\Delta t^{\prime}$ time evolution. 
        The two-qubit unitary gate $\hat{u}$ stays the same as the conventional one shwon in Fig.~\ref{fig:Unitprop}.
        The red dashed rectangles denote the pair of CNOT gates offsetting with each other.
        The areas of each Trotter gate $\exp\left(-{\mathrm i}\hat{A}_{k}\Delta t^{\prime}\right)$ and $\exp\left(-{\mathrm i}\hat{B}_{k}\Delta t^{\prime}\right)$ are separated by the dash-dot lines and highlighted by the characters at the bottom of the figure.
    }
    \label{fig:FINAL}
\end{figure}

To design an efficient circuit implementation of $\exp\left(-{\mathrm{i}}\hat{H}_{\mathrm{eff}}\Delta t^{\prime}\right)$, we further expand $\exp\left(-{\mathrm{i}}\hat{H}_{\mathrm{eff}}\Delta t^{\prime}\right)$ in the following form
\begin{equation}
\begin{split}
    \exp\left(-\mathrm{i}\hat{H}_{\mathrm{eff}}\Delta t^{\prime}\right)
    &=\exp\left(
            -{\mathrm{i}}(\hat{h}^{(1)}+\hat{h}^{(2)})\Delta t^{\prime}/\sqrt{2}
        \right) \\
    &\quad\times \exp\left(
        {\mathrm{i}}(\hat{\sigma}_z^{(1)}\otimes\hat{\sigma}_x^{(2)}+\hat{\sigma}_x^{(1)}\otimes\hat{\sigma}_z^{(2)})\Delta t^{\prime}
        \right) \\
    &\quad\times \exp\left(
        -{\mathrm{i}}(\hat{h}^{(1)}+\hat{h}^{(2)})\Delta t^{\prime}/\sqrt{2}
    \right) \\
    &\quad+\mathcal{O}(m^{-3}), \label{eq:propeff}
\end{split}
\end{equation}
where the third order error $\mathcal{O}(m^{-3})$ is with a higher than that of the conventional Trotter decomposition, derived from the known decomposition
\begin{equation}
\begin{split}
    \exp\left(\frac{\hat{X}+\hat{Y}}{m}\right)
    = \exp\left(\frac{\hat{X}}{2m}\right)
        \exp\left(\frac{\hat{Y}}{m}\right)
        \exp\left(\frac{\hat{X}}{2m}\right)
    + \mathcal{O}(m^{-3}).
\end{split}
\end{equation}
Thus, the proposed construction does not crucially affect on the numerical error for the time propagation by applying sufficient number of Trotter iterations $m$.
This provides a quantum circuit of the time evolution $\exp\left(-{\mathrm{i}}\hat{H}_{\mathrm{eff}}\Delta t^{\prime}\right)$, represented in Fig.~\ref{fig:Trotter_unit}.
The number of required CNOT gates amounts up to eight for implementing $\exp\left(-{\mathrm i}\hat{H}_{\rm eff}\Delta t^{\prime}\right)$ as a quantum circuit.

\begin{figure*}[htbp]
    \centering
    \subfloat[]{
        \includegraphics[width=0.5\linewidth]{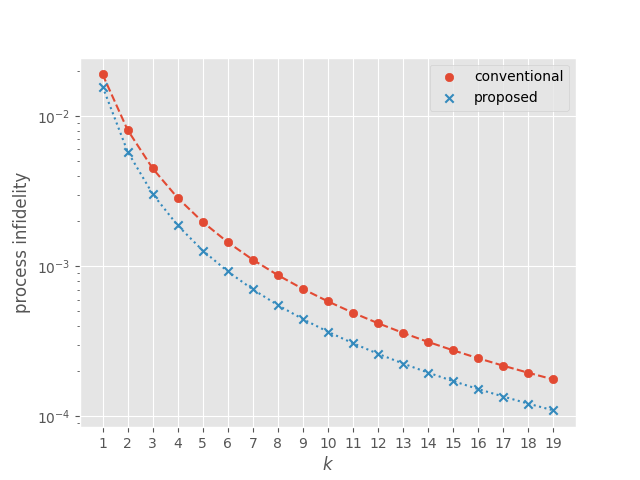}
    }
    % \hfill
    \subfloat[]{
        \includegraphics[width=0.5\linewidth]{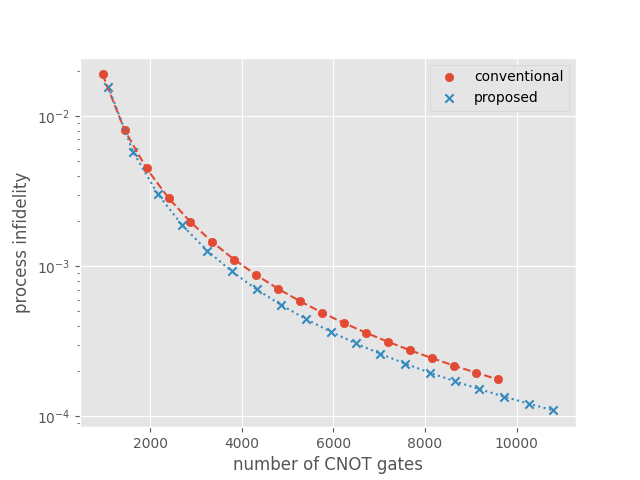}
    }
    \caption{
        The process infidelity to the theoretically predicted time evolution without noise.
        Plots of the conventional Trotter decomposition (Fig.~\ref{fig:Unitprop}) are colored red and those of the proposed decomposition (Fig.~\ref{fig:FINAL}) are colored blue.
        (a) The process infidelity of the evolved state to the ideally evolved state without noise, scaling with the number of Trotter iterations.
        The x-axis $k$ denotes the $k$-th iteration of $n\in\{40, 60, 80, \ldots, 400\}$ for the conventional method and $m\in\{20, 30, 40, \ldots, 200\}$ for the proposed method to make the time intervals $\Delta t$ and $\Delta t^{\prime}$ consistent between each method, i.e. $\Delta t = \Delta t^{\prime}$.
        (b) The process infidelity of the evolved state to the ideally evolved state without noise, scaling with the number of CNOT gates.
        }
    \label{fig:infidelity_process_9-qubit}
\end{figure*}

Using the circuit implementation of $\exp\left(-{\mathrm{i}}\hat{A}_{k}\Delta t^{\prime}\right)$ and $\exp\left(-{\mathrm{i}}\hat{B}_{k}\Delta t^{\prime}\right)$, the whole quantum circuit of the proposed Trotter block for $\Delta t^{\prime}$ in Eq.~\eqref{eq:New_Trotter} is then described by Fig.~\ref{fig:FINAL}.
The single Trotter block requires $8\times4=32$ CNOT gates per four qubits.
Moreover, by considering the offsets between the two CNOT gates in the red-dashed rectangles in Fig.\ref{fig:FINAL}, we can finally reduce CNOT gates to $28$ in each four-qubit Trotter block.
Thus, our circuit construction consumes on average seven CNOT gates per qubit in a single Trotter block with evolution time $2\Delta t^{\prime}$.
Moreover, taking the ratio between $\Delta t$ and $\Delta t^{\prime}$ into account, our circuit construction requires only $7m=1.75n$ CNOT gates to perform $t=n\Delta t$ time evolution for.
Since the conventional Trotter circuit requires an average of $3n$ CNOT gates per qubit up to time $t$, yielding a reduction rate of $1.75n/3n = 0.583$ compared to the original Trotter blocks.
This reduction of CNOT gates significantly contributes to the noise resilience of our proposed Trotter decomposition.

\section{\label{sec:experiments}Experiments}

To demonstrate the practicality of the proposed Trotter decomposition, we simulate the time evolution of the $XXX$ Heisenberg model and calculate the fidelity $F(\tilde{\rho}, \rho_{\mathrm{ideal}})=\operatorname{Tr} \left[\left(\rho_{\mathrm{ideal}}^{1/2} \tilde{\rho} \rho_{\mathrm{ideal}}^{1/2}\right)^{1/2}\right]^{2}$ of the resulting state $\tilde{\rho}$ to the ideally evolved state $\rho_{\mathrm{ideal}}$.

We also compare the process fidelity to the accurate time evolution from $t=0$ to $t=\pi$ between the conventional Trotter unitary operation and the proposed Trotter unitary operation.
Note that the process fidelity\cite{nielsen00} is defined as the state fidelity between Choi matrices of the two completely positive trace-preserving (CPTP) maps of interest.
In the following experiments, we use the noisy and noise-free density matrix simulator, real-device emulator \verb+fake_jakarta+, and real quantum device \verb+ibmq_jakarta+ provided by IBM Quantum Platform.

\subsection{Noise-free process fidelity\label{sec:Noise-free_process_fidelity}}

We first examine the advantage in the convergence rate of approximation error of the proposed Trotter decompositions over the conventional one without the initial state dependency.
To see this, we compare the process fidelity of the conventional and of the proposed approaches to the theoretically predicted time evolution without noise.
Here, we use nine-site $XXX$ Heisenberg model up to a fixed evolution time $t=\pi$.
For the conventional Trotter circuit (Fig.~\ref{fig:Unitprop}) we perform $19$ different Trotter iterations among $n\in\{40, 60, 80, \ldots, 400\}$, and for the proposed Trotter circuit (Fig.~\ref{fig:FINAL}), among $m\in\{20, 30, 40, \ldots, 200\}$.

The results are shown in Fig.~\ref{fig:infidelity_process_9-qubit}, where we plot the process \emph{infidelity} of the conventional method and the proposed method.
Figure~\ref{fig:infidelity_process_9-qubit}(a) compares the process infidelity with different Trotter iterations where both methods use the same time intervals $\Delta t=\Delta t^{\prime}$.
Figure~\ref{fig:infidelity_process_9-qubit}(b) compares the process infidelity on the same basis of the number of the total CNOT gates in the Trotter circuit.
We observe that the process infidelity of the proposed method seems to be halved from that of the conventional one in Fig.~\ref{fig:infidelity_process_9-qubit}(a), which is consistent with theoretical prediction under the setup of $\Delta t=\Delta t^{\prime}$ and $m=n/2$.
Besides, Fig.~\ref{fig:infidelity_process_9-qubit}(b) implies that the proposed method is more noise-robust in the sense that it uses fewer CNOT gates which are considered to be the main source of noise on the current quantum hardware.

\subsection{State fidelity under depolarizing noise\label{sec:State_fidelity_under_depolarizing_noise}}

\begin{figure*}[htbp]
    \centering
    \subfloat[]{
        \includegraphics[width=0.5\linewidth]{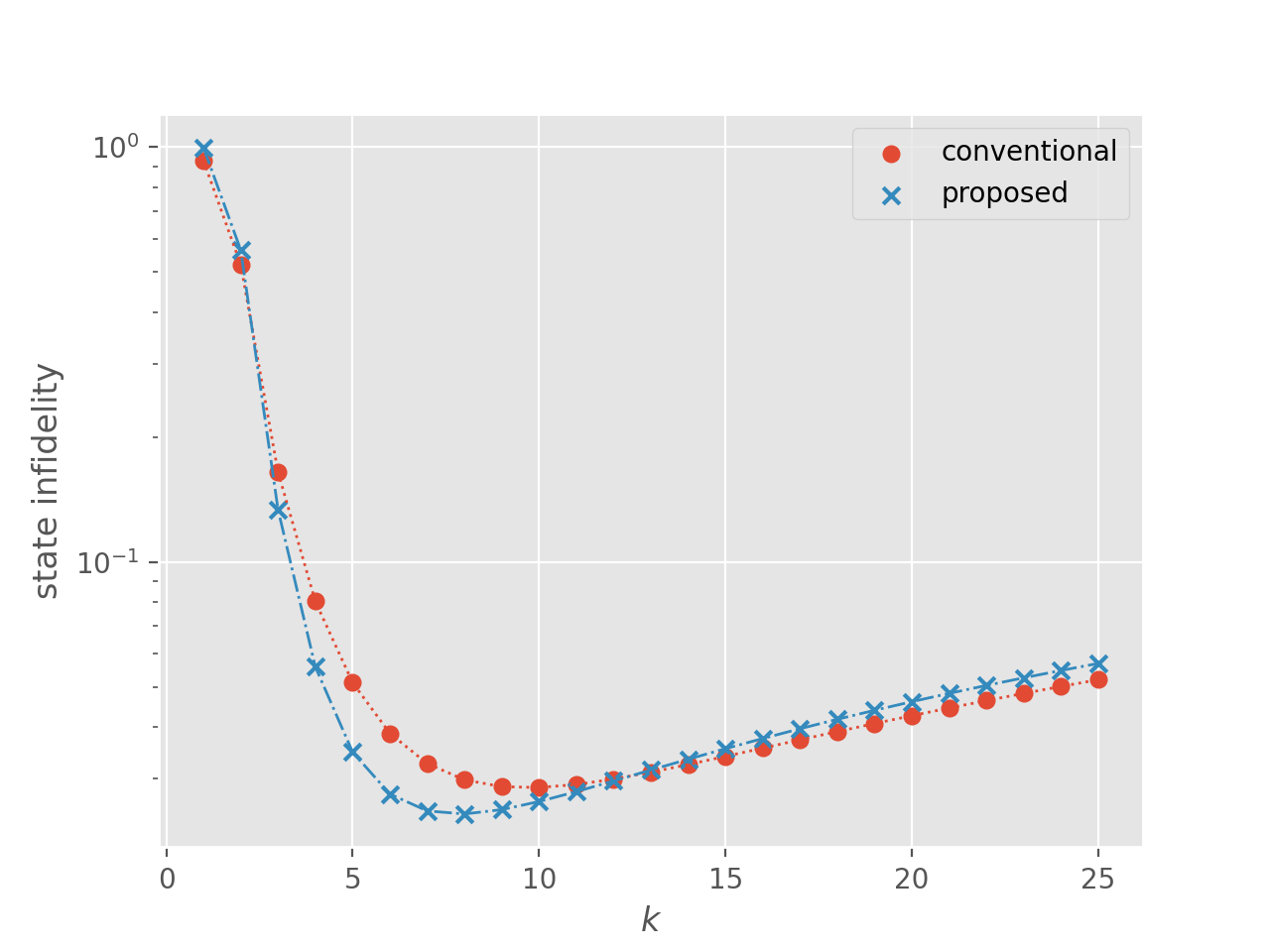}
    }
    % \hfill
    \subfloat[]{
        \includegraphics[width=0.5\linewidth]{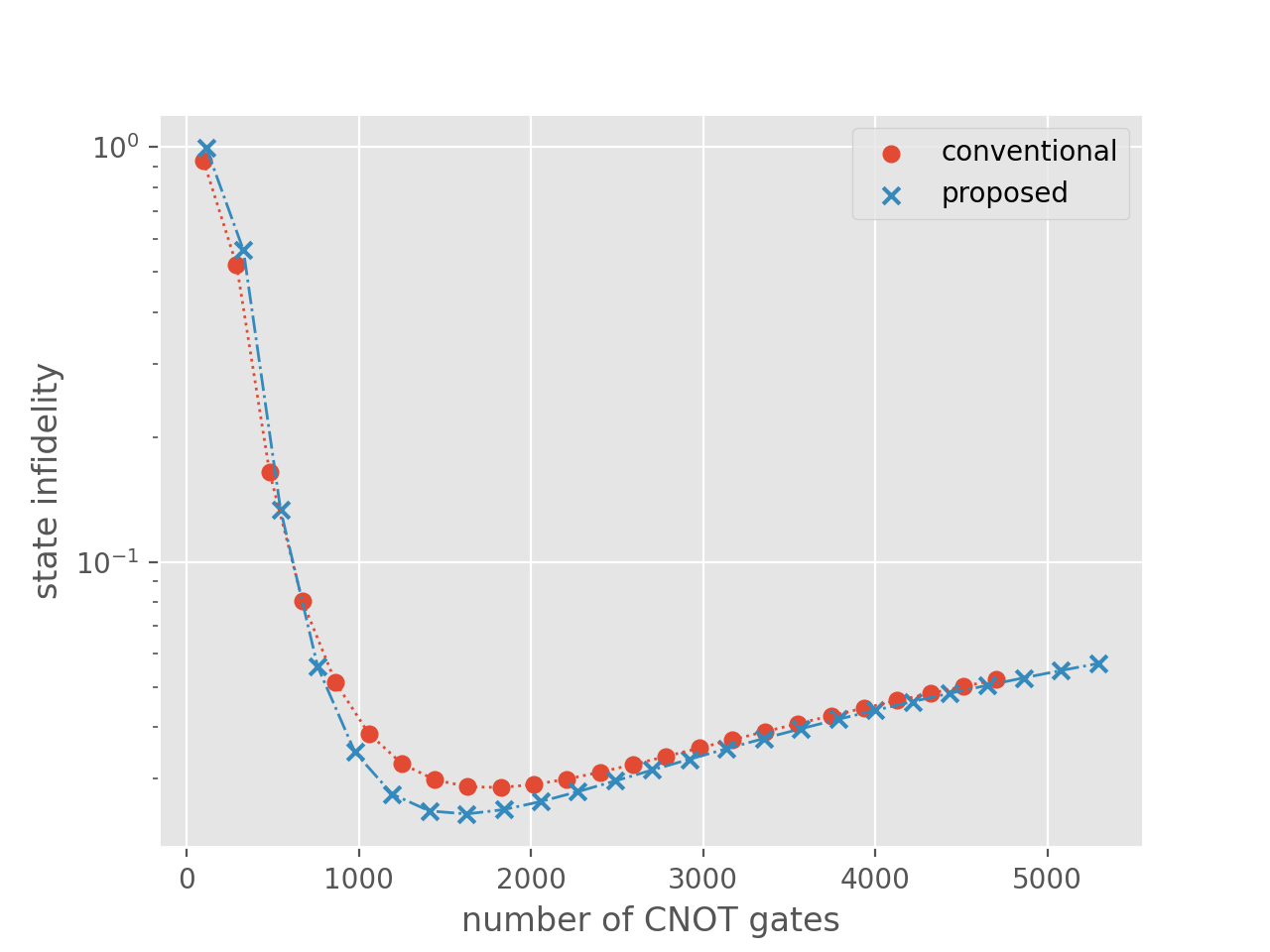}
    }
    \caption{
        The state infidelity of noisy time evolution starting from the Néel state $|101010101\rangle$ to the ideally evolved state.
        Plots of the conventional Trotter decomposition (Fig.~\ref{fig:Unitprop}) are colored red and those of the proposed decomposition (Fig.~\ref{fig:FINAL}) are colored blue.
        (a) The state infidelity of the evolved state under the depolarizing noise with $p_{1}=1.0\times10^{-6}$ to the ideally evolved state, scaling with the number of Trotter iterations.
        The x-axis $k$ denotes the $k$-th iteration of $n\in\{4, 12, 20, \ldots, 196\}$ for the conventional method and $m\in\{2, 6, 10, \ldots, 98\}$ for the proposed method to make the time intervals $\Delta t$ and $\Delta t^{\prime}$ consistent between each method, i.e. $\Delta t = \Delta t^{\prime}$.
        (b) The state infidelity of the evolved state under the depolarizing noise with $p_{1}=1.0\times10^{-6}$ to the ideally evolved state, scaling with the number of CNOT gates.
    }
    \label{fig:infidelity_9-qubit}
\end{figure*}

\begin{figure}[htbp]
    \centering
    \includegraphics[width=\linewidth]{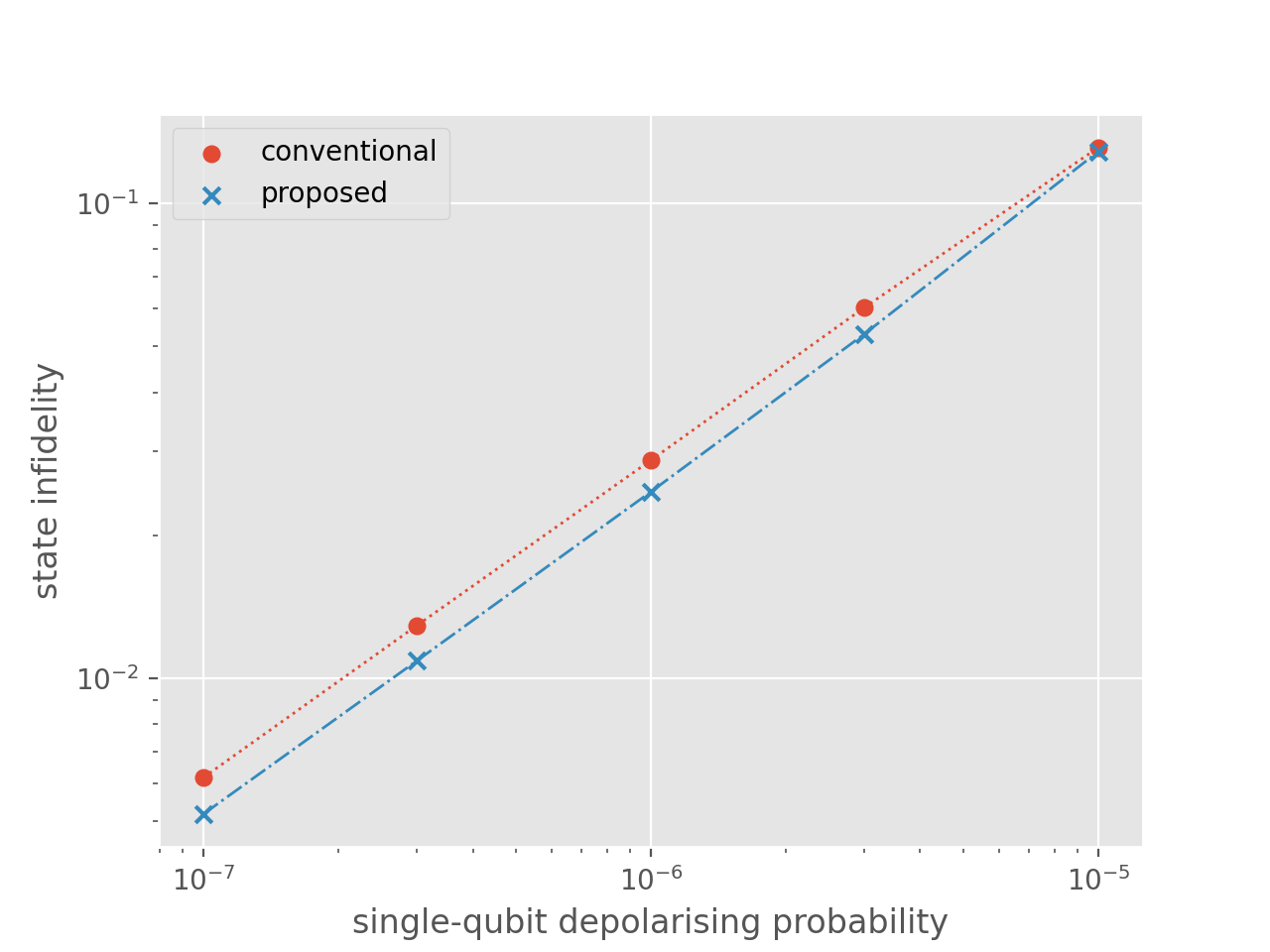}
    \caption{
            The minimum state infidelity of the evolved state under the noise with the conventional and proposed approaches to the ideally evolved state without noise, scaling with different depolarizing probabilities $p_{1}\in\{1.0\times10^{-7},3.0\times10^{-7},1.0\times10^{-6},3.0\times10^{-6},1.0\times10^{-5}\}$ and $p_{2} = 10p_{1}$.
    }
    \label{fig:infidelity_per_noise-level_9-qubit}
\end{figure}

We next perform the noisy numerical simulation of the nine-site $XXX$ Heisenberg model up to a fixed evolution time $t=\pi$.

We introduce depolarizing noise in both single-qubit gates and two-qubit gates with depolarizing probabilities $p_{1} = 1.0\times 10^{-6}$ and $p_{2} = 10p_{1} = 1.0\times 10^{-5}$, respectively, which reflect the noise levels of the current and near-future quantum hardware~\cite{Hughes2025Trapped-ion, Ransford2025Helios}.
Starting from the Néel state $|101010101\rangle$, we compare the state fidelity between the conventional approach for $25$ different Trotter iterations among $n\in\{4, 12, 20, \ldots, 196\}$ and the proposed approach for $25$ different Trotter iterations among $m\in\{2, 6, 10, \ldots, 98\}$.

The simulated results between the conventional Trotter decomposition and the proposed decomposition are plotted in Fig.~\ref{fig:infidelity_9-qubit}.
We observe that there exists an optimal number of Trotter iterations that balances the simulation accuracy and the noise effect induced when increasing the Trotter iterations.
From Fig.~\ref{fig:infidelity_9-qubit}(a), we see that the optimal state infidelity by the proposed method achieves lower state infidelity than the conventional approach.

Note that the cross of the plots between the conventional method and the proposed method when $k=12$ in Fig.~\ref{fig:infidelity_9-qubit}(a) results from the condition to make $\Delta t=\Delta t^{\prime}$ in the x-axis, where the extent of the residual error by the proposed method is designed to be the half of the residual error by the conventional method.
This means that the proposed method uses correspondingly deeper circuits with more CNOT gates to achieve such a better residual error.
Under the noisy execution, this appears as the cross of the plots.
When adjusting the x-axis relative to the number of CNOT gates in Fig.~\ref{fig:infidelity_9-qubit}(b), we then see that the performance of the proposed method is better than the conventional one for larger Trotter iterations.
This implies that our proposed decomposition is more noise-robust to achieve a higher state infidelity under the noisy execution.
We here also remark that the conventional method outperforms the proposed method for smaller Trotter iterations, which can be explained by insufficient precision to approximate $\exp\left(-\mathrm{i}\hat{H}_{\mathrm{eff}}\Delta t^{\prime}\right)$ in Eq.~\eqref{eq:propeff}.
% Since the proposed method larger required $7m=3.5n$ CNOT gates per a qubit in the proposed method than $7m=3.5n$ CNOT gates per a qubit in the conventional one where both methods use the same time intervals $\Delta t=\Delta t^{\prime}$.

Focusing on the infidelity with the optimal number of Trotter iterations when simulating time evolution starting from the Néel state, Fig.~\ref{fig:infidelity_per_noise-level_9-qubit} visualizes the advantage of our proposed method over the conventional method in terms of lower infidelity under different noise levels ranging among $p_{1}\in\{1.0\times10^{-7},3.0\times10^{-7},1.0\times10^{-6},3.0\times10^{-6},1.0\times10^{-5}\}$ and $p_{2} = 10p_{1}$.
Note that the latest quantum devices have already reach the regime where the single-qubit and two-qubit gate error probability take $10^{-5}$ and $10^{-4}$, respectively~\cite{Hughes2025Trapped-ion, Ransford2025Helios}.
Observing from Fig.~\ref{fig:infidelity_per_noise-level_9-qubit}, the advantage of the proposed method becomes clearer when the error rates become smaller.
This means that our approach would further outperform the conventional approach in the near-future quantum devices with smaller noise levels.

\begin{figure*}[htbp]
    \centering
    \subfloat[]{
        \includegraphics[width=0.5\linewidth]{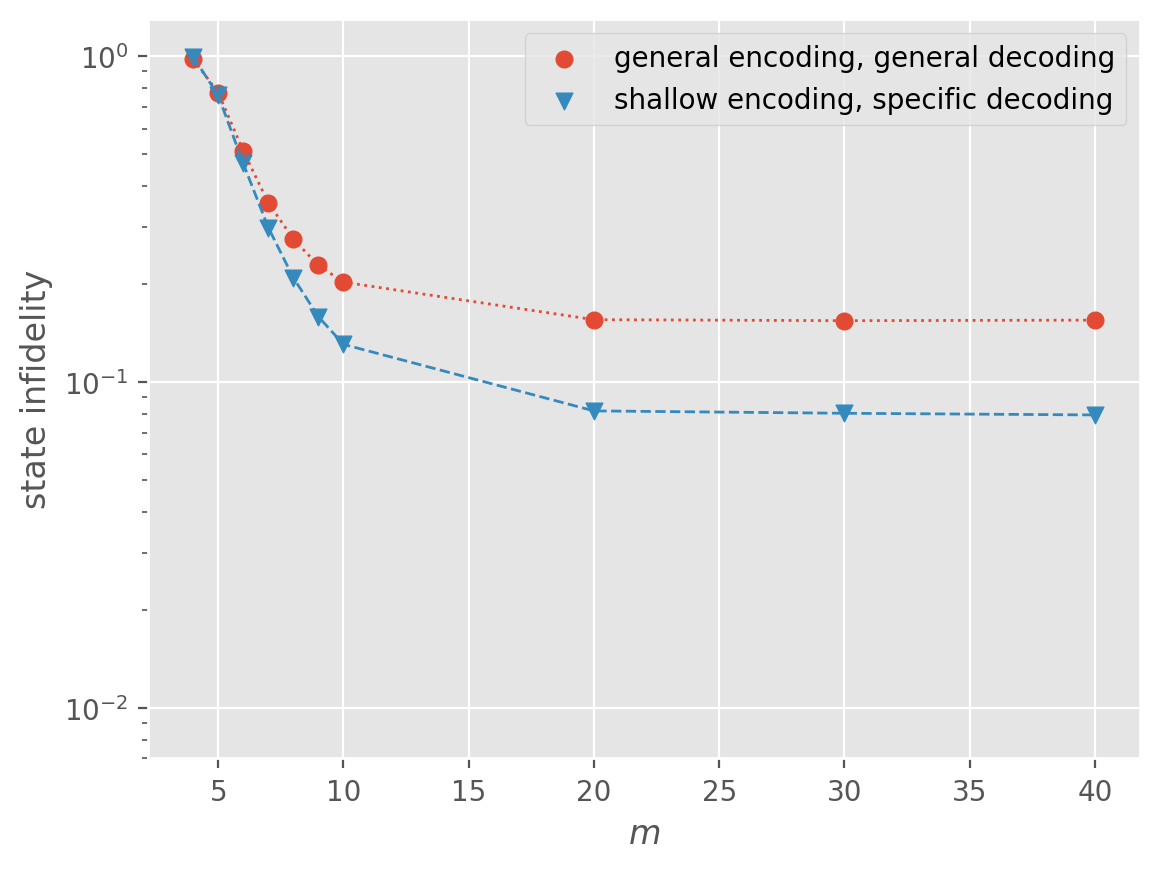}
    }
    % \hfill
    \subfloat[]{
        \includegraphics[width=0.5\linewidth]{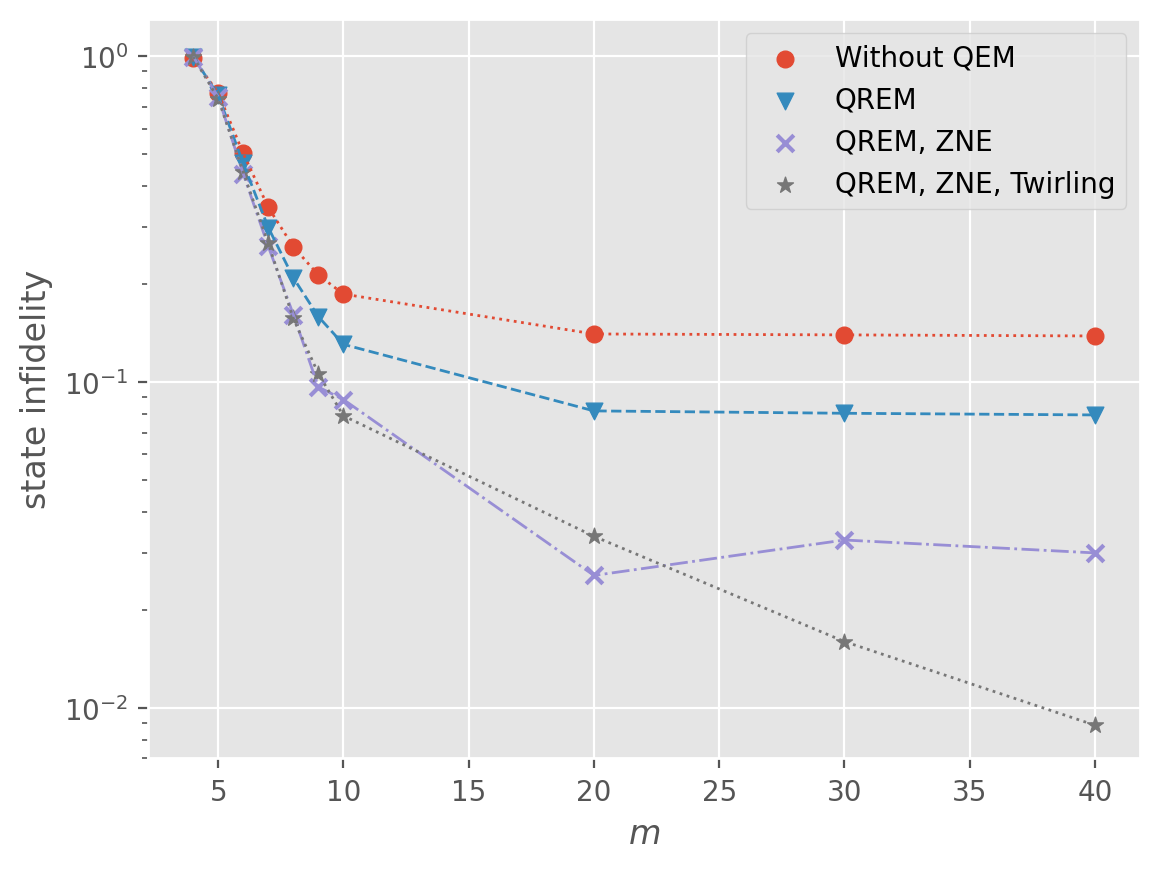}
    }
    \caption{
        (a) The state infidelity of the evolved state to the expected noise-free state, simulated by the two encoding-decoding strategies among different Trotter iterations $m$.
        (b) The state infidelity of the evolved state to the expected noise-free state, for different QEM levels among different Trotter iterations $m$.
    }
    \label{fig:result_ibmq_jakarta}
\end{figure*}

\subsection{Real-device experiments with error suppression\label{sec:Real-device_experiments_with_error_suppression}}

\begin{figure}[htbp]
    \centering
    \includegraphics[width=\linewidth]{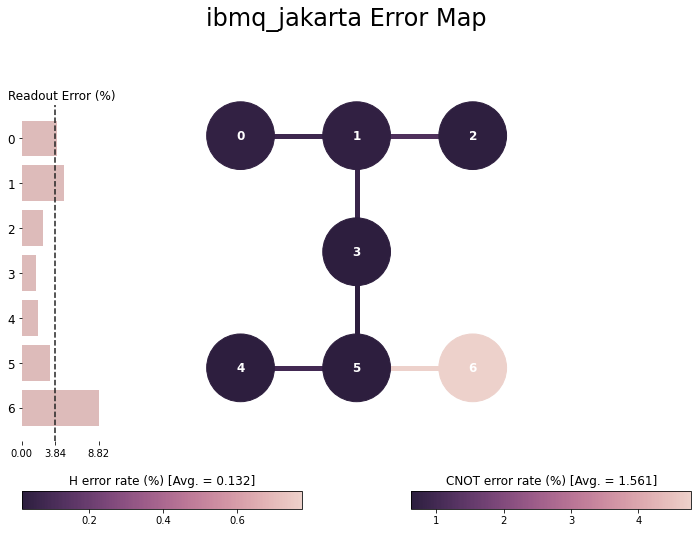}
    \caption{\label{fig:ibmq_jakarta}
        The error map of ibmq\_jakarta on April 16, 2022.
        The numbers on the figure represent the indices of physical qubits.
        We use the physical qubits 5, 3, and 1 with the virtual qubit indices 0, 1, and 2 on quantum circuits.
        The device noise is subject to temporal fluctuations.
    }
\end{figure}

In implementing the proposed Trotter decomposition on real quantum hardware \verb+ibmq_jakarta+ and its noise-calibrated simulator \verb+fake_jakarta+, we simulate the time evolution of the three-site $XXX$ Heisenberg model from $t=0$ to $t=\pi$.
This real device experiments with $N=3$ corresponds to using only the propagator $\exp\left(-{\mathrm i}\hat{A}_{1}\Delta t^{\prime}\right)$ in Eq.~\eqref{eq:New_Trotter_error}, which can be seen as the simplest case of the proposed method.
% In this implementation with $N=3$, we take the propagator $\exp\left(-{\mathrm i}\hat{A}_{1}\Delta t^{\prime}\right)$ only and drop out the propagators $\exp\left(-{\mathrm i}\hat{B}_{2}\Delta t^{\prime}\right)\exp\left(-{\mathrm i}\hat{A}_{2}\Delta t^{\prime}\right)$ in Eq.~\eqref{eq:New_Trotter}.}

Since \verb+ibmq_jakarta+ has constrained qubit connectivity shown in Fig.~\ref{fig:ibmq_jakarta}, we further reduce the circuit depth and the number of CNOT gates by adopting the ``shallow'' encoding and ``specific'' decoding methods, in which the encoding and decoding processes are simplified regarding the subspace that the chosen initial state belongs to.
For example, the encoding operation $\hat{U}_{\mathrm{enc}}$ transforms the initial state $\ket{110}$ into $\ket{010}$, which can be equivalently realized by applying $\hat{\sigma}_x^{(1)}\otimes\hat{1}^{(2)}\otimes\hat{1}^{(3)}$ to the initial state $\ket{110}$.
Besides, given an initial state within the subspace of $P=1$, which is the case for $\ket{110}$, the evolved state at any evolution time $t$ would ideally stay in the same subspace. 
This allows us to further reduce the CNOT operations in the decoding process: acting CNOT(2$\rightarrow$1) and CNOT(3$\rightarrow$2) on the obtained final state sequentially. 
This optimized encoding-decoding process makes the proposed decomposition more compatible with near-term superconducting devices, without changing the targeted physical evolution.

Based on the above, we compute the infidelity of the resulting state at the evolution time $t = \pi$ evolved from a given initial state $|110\rangle$ at $t=0$.
We add quantum error mitigation (QEM)~\cite{li2017efficient, Temme_2017, endo2018practical, digital_zne, yang2022efficient, Yang2025Resource-Efficient, cai2023quantum, endo2021hybrid} to reduce the noise effect through classical post-processing.
Particularly, we use quantum readout error mitigation (QREM)~\cite{yang2022efficient} and zero-noise extrapolation (ZNE)~\cite{Temme_2017, digital_zne}.
We use the digital ZNE method~\cite{digital_zne} with the linear fitting method and the scale factors $1.0$, $2.0$ and $3.0$, provided by Mitiq~\cite{mitiq}.
Furthermore, the Pauli twirling technique~\cite{Bennett1996Purification, Knill2004Fault-Tolerant, Wallman2016Noise} is also combined with ZNE, referring to the implementation by Berthusen et al.~\cite{Berthusen_2021}.
Each quantum circuit is executed with $8192$ shots, and the infidelity is averaged over $8$ samples.

First, we examine the performance among Trotter iterations $\{4,5,6,7,8,9,10,20,30,40\}$ on the noisy simulator \verb+fake_jakarta+, comparing the two encoding-decoding methodologies: the general encoder $\hat{U}_{\mathrm{enc}}$ and general decoder $\hat{U}_{\mathrm{enc}}^{\dagger}$ (general-general), and the aforementioned shallow encoding and specific encoding (shallow-specific).
Here, we apply only QREM to noisy results before computing the infidelity.
The result is shown in Fig.~\ref{fig:result_ibmq_jakarta}(a), where both encoding-decoding methods achieve the infidelity below $0.2$ with more than $10$ Trotter iterations and the shallow-specific method further achieves the infidelity smaller than $0.1$.

The effect of QEM methods is also investigated under \verb+fake_jakarta+.
Here, we set the configuration to shallow encoding and specific decoding.
We observe from Fig.~\ref{fig:result_ibmq_jakarta}(b) that both QREM and ZNE contribute to reducing the infidelity.
The instability of ZNE can be improved by adding Pauli twirling that tailors the noise to a stochastic Pauli channel.
We also see that combining Pauli twirling further enhances the accuracy gains achieved by finer Trotter decomposition.

\begin{table}[htbp]
    \centering
    \begin{tabular}{lcc} \hline \hline
        Settings & \verb+fake_jakarta+ & \verb+ibmq_jakarta+ \\ \hline 
        {\bf General-general}\\
        Without QEM & $0.7856 \pm 0.0015$ & $0.8039 \pm 0.0048$ \\ 
        QREM & $0.8448 \pm 0.0015$ & $0.9032 \pm 0.0054$ \\ 
        QREM, ZNE & $0.9393 \pm 0.0053$ & $0.9866 \pm 0.0017$ \\ 
        QREM, ZNE, Twirling & $0.9801 \pm 0.0031$ & - \\ 
        {\bf Shallow-specific}\\
        Without QEM & $0.8631 \pm 0.0017$ & $0.8637 \pm 0.0041$ \\ 
        QREM & $0.9234 \pm 0.0016$ & $0.9728 \pm 0.0040$ \\ 
        QREM, ZNE & $0.9840 \pm 0.0024$ & $0.9857 \pm 0.0043$ \\ 
        QREM, ZNE, Twirling & $0.9714 \pm 0.0048$ & $0.9624 \pm 0.0167$ \\ 
        \hline \hline
    \end{tabular}
    \caption{
        The fidelity of the simulated state from IBM Quantum Jakarta and its fake simulator under different QREM levels and encoding-decoding strategies.
        ``General-general'' represents the use of the general encoder and the general decoder, and ``Shallow-specific'' represents the use of the shallow encoder and the specific decoder.
    }
    \label{tab:ibmq_jakarta}
\end{table}

Finally, we examine the time evolution from $t=0$ to $t=\pi$ on the real quantum device \verb+ibmq_jakarta+.
We execute $100$ Trotter iterations with and without QEM under the two encoding-decoding methods.
The state fidelity is then calculated by reconstructing the density matrix through state tomography~\cite{Raymer1994Complex, Leonhardt1995Quantum-State, Leibfried1996Experimental}.

The results are listed in Table~\ref{tab:ibmq_jakarta}, where the fidelities obtained by \verb+fake_jakarta+ and \verb+ibmq_jakarta+ are compared.
All the fidelities exceed $0.80$ on \verb+ibmq_jakarta+, and all the fidelities even exceed $0.90$ with QREM only.
Remarkably, we achieve a fidelity over $0.98$ with ZNE with the general-general method, which ensures the generality of our method in simulating the dynamics from an arbitrary initial state.
All the experimental results on the noisy simulator and the real device support the practicality of our proposed Trotter decomposition.

\section{\label{sec:conclusion}Conclusion}

In this work, we propose a novel, noise-resilient Trotter decomposition focusing on the symmetry of the given Heisenberg Hamiltonian to notably reduce the number of CNOT gates in its circuit implementation without sacrificing the accuracy of the Trotter decomposition.
The noise-robustness of the proposed method is demonstrated through the numerical simulation under noise.
Our experiments also record high fidelities in simulating the three-site Heisenberg model on the real quantum device.
The proposed method thus establishes a direct link between physical insight into the model's symmetry and quantum circuit design regarding efficient Trotter decomposition, offering both a fresh theoretical perspective and practical benefits for performing noise-resilient quantum dynamics simulation.

This work opens up several promising directions.
First, owing to the generality of the proposed approach, it can be extended to a broader class of physical and chemical models.
In particular, our framework applies to models that share the same underlying symmetric structure as the Heisenberg model after an equivalent mapping, such as the Jordan-Wigner transformation~\cite{JWT1,JWT2}, which maps fermionic creation and annihilation operators to Pauli operators.
Through such transformations, electronic quantum states would be able be simulated within our present framework.
A systematic characterization of the classes of physical systems that can be equivalently mapped onto this symmetric form is left as a significant direction for future work.

Next, integrating more recent QEM techniques based on ZNE~\cite{Endo2019Mitigating,Hakkaku2025Data-Efficient} would further suppress both coherent and algorithmic errors in Trotterized circuits.
Moreover, resource-efficient implementations using subspace expansion methods~\cite{Yoshioka2022Generalized,Yang2025Resource-Efficient} can also be adapted to our framework to mitigate both coherent and stochastic errors.
In combination with quantum–classical divide-and-conquer approaches~\cite{Sun2022Perturbative,Eddins2022Doubling,Yuan2021Quantum,Harada2025densitymatrix}, these techniques would further enhance the practical feasibility of our method on near-term quantum devices.

\begin{acknowledgments}
The result of real-device experiments is obtained as a solution to the IBM Quantum Awards: Open Science Prize 2021, for which the initial manuscript has been publicly available at~\cite{github_osp_solutions} since April 2022.
B.Y. and N.N. sincerely thank all those who made this contest possible.
\end{acknowledgments}

\section*{Author Declarations}

\subsection*{Conflict of Interest}
The authors have no conflicts to disclose.

\subsection*{Author Contributions}
\textbf{Bo Yang:}
Conceptualization (equal);
Data curation (lead);
Formal analysis (supporting);
Funding acquisition (equal);
Investigation (lead);
Methodology (supporting);
Project administration (equal);
Resources (equal);
Software (lead);
Supervision (equal);
Validation (equal);
Visualization (equal);
Writing - original draft (equal);
Writing - review \& editing (equal).

\textbf{Naoki Negishi:}
Conceptualization (equal);
Data curation (supporting);
Formal analysis (lead);
Funding acquisition (equal);
Investigation (supporting);
Methodology (lead);
Project administration (equal);
Resources (equal);
Software (supporting);
Supervision (equal);
Validation (equal);
Visualization (equal);
Writing - original draft (equal);
Writing - review \& editing (equal).

\section*{Data Availability Statement}
The code and data used in the experiments in this work are publicly available on the GitHub page: https://github.com/BOBO1997/osp\_solutions~\cite{github_osp_solutions}.

\bibliography{main}

\end{document}